\documentclass[pdflatex,sn-mathphys]{sn-jnl}

\jyear{2021}%

\theoremstyle{thmstyleone}%
\newcommand{\framework}{atomic parallelism}
\newcommand{\newabs}{segment group}
\newcommand{\Framework}{Atomic parallelism}
\newcommand{\Newabs}{Segment group}
\theoremstyle{thmstyletwo}%

\theoremstyle{thmstylethree}%

\begin{document}

\title[Sgap: Towards Efficient Sparse Tensor Algebra Compilation for GPU]{Sgap: Towards Efficient Sparse Tensor Algebra Compilation for GPU}


\author[1]{\fnm{Genghan} \sur{Zhang}}
\author[1]{\fnm{Yuetong} \sur{Zhao}}
\author[1]{\fnm{Yanting} \sur{Tao}}
\author[2]{\fnm{Zhongming} \sur{Yu}}
\author*[3]{\fnm{Guohao} \sur{Dai}}\email{daiguohao@sjtu.edu.cn}
\author[4]{\fnm{Sitao} \sur{Huang}}
\author[5]{\fnm{Yuan} \sur{Wen}}
\author[6]{\fnm{Pavlos} \sur{Petoumenos}}
\author*[1]{\fnm{Yu} \sur{Wang}}\email{yu-wang@tsinghua.edu.cn}
\affil[1]{\orgdiv{Department of Electronic Engineering}, \orgname{Tsinghua University}, \orgaddress{\street{Rhom 4101}, \city{Beijing}, \postcode{100084}, \country{China}}}
\affil[2]{\orgdiv{Department of Computer Science and Enigeering}, \orgname{University of California San Diego}, \orgaddress{\street{Gilman Drive}, \city{La Jolla}, \postcode{92093}, \state{CA}, \country{USA}}}
\affil[3]{\orgdiv{Qingyuan Research Institute}, \orgname{Shanghai Jiao Tong University}, \orgaddress{\street{Room 318A, Building A No. 930 Jianchuan Road}, \city{Shanghai}, \postcode{200240}, \country{China}}}
\affil[4]{\orgdiv{Department of Electrical Engineering and Computer Science}, \orgname{University of California Irvine}, \orgaddress{\street{3215 Engineering Hall}, \city{Irvine}, \postcode{92697}, \state{CA}, \country{USA}}}
\affil[5]{\orgdiv{Department of Computer Science}, \orgname{University of Aberdeen King's College},\orgaddress{Meston Building}, 
\city{Aberdeen},\postcode{AB24 3UE}, \country{United Kingdom}}
\affil[6]{\orgdiv{Department of Computer Science}, \orgname{University of Manchester}, \orgaddress{Kilburn Building}, \city{Manchester}, \postcode{M13 9PL}, \country{United Kingdom}}

\abstract{Sparse compiler is a promising solution for sparse tensor algebra optimization. In compiler implementation, \textbf{reduction} in sparse-dense hybrid algebra plays a key role in performance. Though GPU provides various reduction semantics that can better utilize the parallel computing and memory bandwidth capacity, the central question is: \textit{how to elevate the \textbf{flexible reduction semantics} to sparse compilation theory that assumes serial execution}. 
Specifically, we have to tackle two main challenges: (1) there are wasted parallelism by adopting static synchronization granularity (2) static reduction strategy limits optimization space exploration. We propose Sgap: \textbf{\textit{\underline{s}egment \underline{g}roup}} and \textbf{\textit{\underline{a}tomic \underline{p}arallelism}} to solve these problems. {\Framework} captures the flexible reduction semantics to systematically analyze the optimization space of sparse-dense hybrid algebra on GPU. It is a new optimization technique beyond current compiler-based and open-source runtime libraries. {\Newabs} elevates the flexible reduction semantics to suitable levels of abstraction in the sparse compilation theory. It adopts changeable group size and user-defined reduction strategy to solve challenge (1) and (2), respectively. Finally, we use GPU sparse matrix-matrix multiplication (SpMM) on the TACO compiler as a use case to demonstrate the effectiveness of {\newabs} in reduction semantics elevation. We achieve up to $1.2\times$ speedup over the original TACO's SpMM kernels. We also apply new optimization techniques found by {\framework} to an open-source state-of-the-art SpMM library dgSPARSE. We achieve $1.6\times\sim2.3\times$ speedup on the algorithm tuned with {\framework}.}




\maketitle

\section{Introduction}
Sparse tensor algebra has been widely used in many fields, including machine learning~\cite{hamilton2017inductive,kipf2016semi,liu2015sparse}, data analysis~\cite{kolda2009tensor}, scientific computing~\cite{shantharam2011characterizing,bell2012exposing}, graph processing~\cite{yuster2004detecting}. However, it is challenging to optimize sparse tensor applications because of diversity in computation patterns and irregularity in memory access behavior. Sparse compilers have shown great potential to solve this problem. Sparse compilers can use \textbf{one} monolithic theory to express diverse data formats and operations, and provide flexible user interface, enabling users to explore the optimization space given data and hardware. Therefore, more and more researchers are turning to sparse compilers for general solutions~\cite{bik1993compilation,venkat2015loop,strout2018sparse,kjolstad:2017:taco,kjolstad:2020:phd-thesis,popoola2021spf,bik2022compiler,ye2022sparsetir}.

 However, it is challenging to design a sparse compiler that can both compile various algebras and generate highly optimized code. In particular, \textit{sparse-dense hybrid algebra} on GPU brings unique challenges to sparse compilers. After analysing sparse-dense hybrid algebra's mathematical expression, we find out that \textbf{reduction} is its key operation~\cite{nisa2019mttkrp,huang2020ge,kurt2022ttm}. There are several possible ways to do reduction on GPUs. Different reduction methods are preferred for different workloads. Choosing the correct reduction method can accelerate kernels~\cite{dai2022heuristic,bell2009implementing}. For example, controlled experiments in~\cite{dai2022heuristic} show that parallel reduction can outperform conditional reduction and vice versa by $2\times\sim4\times$. However, current sparse compilers lack the abstraction for such flexible reduction semantics. That is because they assume the code executes serially. GPU reduction is different from the serial reduction in that it changes the reduction code's structure (e.g., control-flow and loop basic block). Therefore, it cannot be naively generated by directly adding or replacing some instructions like the \textit{unroll} in CPU. Solving this problem requires elevating reduction semantics to the sparse compilation theory in a systematic way. 
 
\begin{figure}[h]%
\centering
\includegraphics[width=0.9\textwidth]{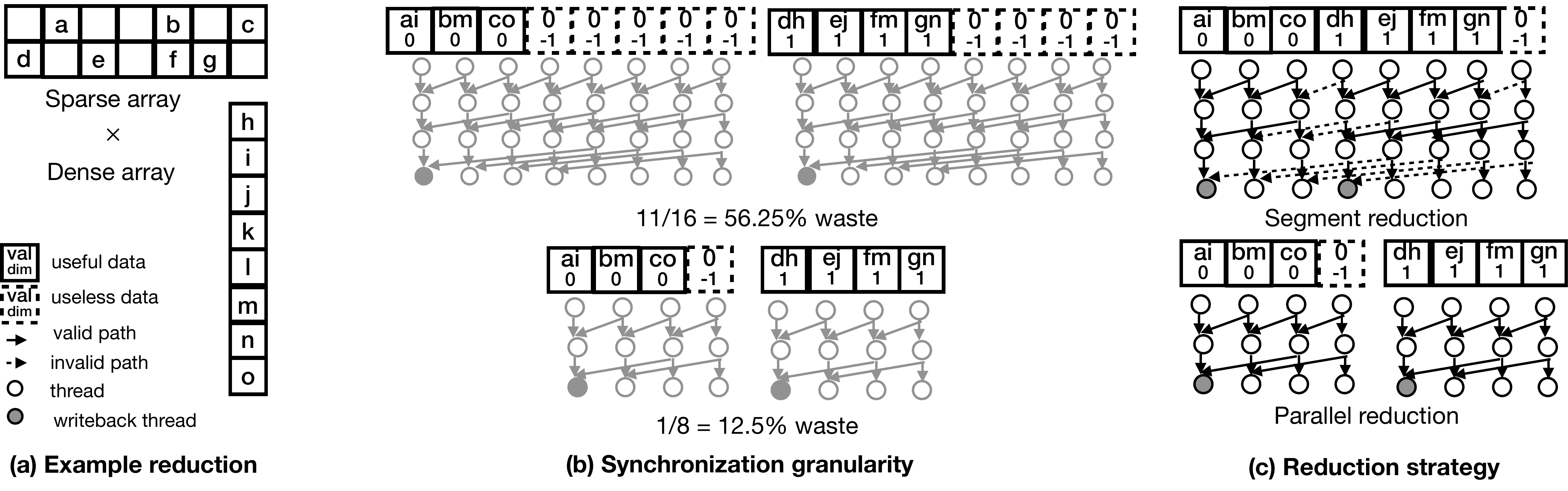}
\caption{Sparse compilers suffer from static synchronization granularity and static reduction strategy. (a) Example reduction with
legends in latter subfigures. (b) Parallelism waste caused by improper synchronization granularity. (c) One type of segment reduction and one type of parallel reduction. Segment reduction has two writeback threads and parallel reduction has one.}\label{fig:intro}
\end{figure}
 
 However, elevating the flexible reduction semantics to sparse compilation theory faces two main challenges: (1) \textbf{Static synchronization granularity wastes parallelism}: GPU synchronizes a group of threads whose group size is power of 2, which we term as synchronization granularity. Threads can pass local register values to another thread in the same group. However, static synchronization granularity may waste parallelism when inputs are dynamic. For example, if not all threads' register values are gathered, threads that do not influence the reduction result still have to wait to be synchronized. In other words, the synchronization granularity is too large for such input data, as is shown in Fig.~\ref{fig:intro} (b). However, current sparse compilers only assume synchronization granularity to be 32, which wastes the parallelism. This is the limitation of current sparse compilers. (2) \textbf{Static reduction strategy limits optimization space exploration}: GPU has provided very flexible methods to do reduction. Multiple threads in a thread group will write back to the final results. We name such thread \textit{writeback thread}. There could be more than one writeback thread in a thread group. The thread indices of writeback threads can also be decided at runtime and are controlled by the reduction strategy. Different algorithms favor different reduction strategies. For example, as is shown in Fig.~\ref{fig:intro} (c), if we assign a given number of non-zeros to each thread group, it has to use segment reduction. That is because threads need to write back according to the coordinate and thus writeback thread is decided at runtime. However, in another algorithm where all threads in a group are guaranteed to write back to the same place, it can use parallel reduction~\cite{bell2009implementing}. However, current sparse compilers assume that only the first thread in a thread group is the writeback thread and use parallel reduction. 

\begin{figure}[h]%
\centering
\includegraphics[width=0.7\textwidth]{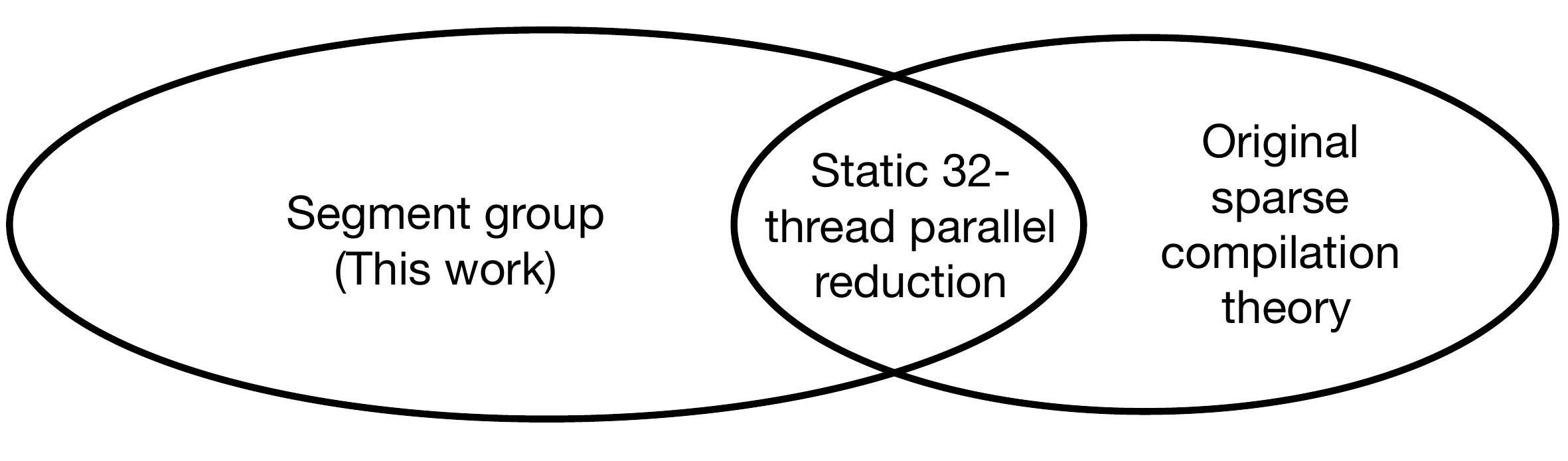}
\caption{Venn diagram for the relation between {\framework} and original sparse compilation theory. The element is the point in the algorithm design space of a sparse-dense hybrid algebra. Original sparse compilation theory can only express parallel reduction with group size 32. However, it can also express some optimization points, for example, loop reorder, beyond {\framework}. The union of segment group and original theory creates a new sparse compilation theory.}\label{fig-set}
\end{figure}

To tackle these challenges and build a more efficient sparse compiler, we propose \textit{\textbf{\framework}} and \textit{\textbf{\newabs}} in this paper and implement our techniques in a real sparse compiler TACO~\cite{kjolstad:2019:workspaces,chou:2018:formats,kjolstad:2017:taco,senanayake:2020:scheduling}. 
{\Framework} models the optimization space of sparse-dense hybrid algebra from the reduction view. It uses the minimal data and reduction parallelism to distinguish different algorithms of a given algebra. Minimal data are used to define reduction strategy and reduction parallelism for synchronization granularity. We use this model to propose new optimization techniques. 
{\Newabs} is a new abstraction for sparse compilation theory. It captures the dynamic synchronization granularity and dynamic reduction strategy. To be specific, we use flexible group size to solve challenge (1) and design full-stack support for user-defined reduction strategy, which solves challenge (2). As is shown in Fig.~\ref{fig-set}, {\newabs} extends the expression ability of original sparse compilation theory.
Finally, we use sparse matrix-matrix multiplication (SpMM) as an example to demonstrate {\framework} and {\newabs}. SpMM is one of the most widely used sparse-dense hybrid algebra. It is the core operator of many emerging applications~\cite{han2016eie,wang2019structured,lin2021es,asgari2021fafnir}. It is also the simplest form of sparse-dense hybrid algebra.

Therefore, this work manages to push the frontier a step forward on these two challenges by a combined method involving \underline{s}egment \underline{g}roup and \underline{a}tomic \underline{p}arallelism which we called \textit{\textbf{Sgap}} in this paper. Our contributions are as follows:

\begin{enumerate}[1.]
\item We propose a framework \textbf{\textit{{\framework}}} to analyse sparse-dense hybrid algebra and propose new SpMM designs beyond previous works~\cite{yang2018design,hong2019adaptive,huang2020ge,mehrabi2021learning,dai2022heuristic}.
\item Based on the {\framework}, we point out that current sparse compilers miss important optimization opportunities. We propose a new abstraction \textbf{\textit{{\newabs}}} for sparse compilers. {\Newabs} can reduce parallelism waste and improve workload balance.
\item We implement {\newabs} in TACO and get up to $1.2\times$ speedup on average over the original TACO's SpMM kernels. Next, we generalize our findings from TACO to dgSPARSE~\cite{dai2022heuristic}, an open-source state-of-the-art SpMM library. We achieve $1.6\times\sim2.3\times$ speedup over dgSPARSE on the algorithm we tune. 
\end{enumerate}
The rest of this paper is organized as follows. Background
information is provided in Section~\ref{sec:bg}. Section~\ref{sec:framework} introduces {\framework} and Section~\ref{sec:newabs} is for {\newabs}. Then the implementation of {\newabs} in TACO is detailed in Section~\ref{sec:newabs4taco}. After that, we illustrate the combination of {\framework} and {\newabs} in TACO. Our evaluation of new SpMM algorithms in TACO and generalization to dgSPARSE is presented in Section~\ref{sec:exp}. The paper is concluded in Section~\ref{sec:conclusion}.

\section{Background}\label{sec:bg}

\subsection{Sparse-dense Hybrid Algebra}
 Sparse-dense hybrid algebra can be defined in two equivalent forms: the tensor formulation (TF) in Eq.~\ref{algebra-view} and the database formulation (DF) in Eq.~\ref{db-view}. From TF sparse-dense hybrid algebra because the operands of it are sparse and dense, for example, MTTKRP (Matricized Tensor Times Khatri Rao Product)~\cite{nisa2019mttkrp}, SDDMM (Sampled Dense-Dense Matrix Multiplication)~\cite{yu2021exploiting}, SpMM (sparse Matrix-Matrix Multiplication)~\cite{huang2020ge}, TTM (Tensor Times Matrix Product)~\cite{kurt2022ttm}.  We use Einstein's summation to define sparse-dense hybrid algebra in AF as Eq.~\ref{algebra-view}.
\begin{equation}\label{algebra-view}
\mathbb{Y}_{y_1, y_2,\cdots,y_M} = \mathbb{A}_{a_1, a_2,\cdots,a_N}\prod_{i=1}^{D}\mathbb{X}^{j}_{x_1^j,x_2^j,\cdot,x_{M^j}^j}
\end{equation}
$\mathbb{Y}$ is the output tensor, $\mathbb{X}^j$ are dense input tensors, and $\mathbb{A}$ is the sparse input tensor. At least one level $a_N$ in $\mathbb{A}$ does not store in dense format. $y_1, y_2,\cdots,y_M$,$a_1, a_2,\cdots,a_N$,$x_1^j,x_2^j,\cdots,x_{M^j}^j$ are in the same index variable set. $M$ is the mode of output tensor, and $N$ is the mode of sparse input tensor. $D$ is the number of dense input tensors, and $M^j$ is the mode of dense input tensor $\mathbb{X}^j$. Specifically, MTTKRP, TTM, SDDMM, and SpMM are expressed as:
\begin{subequations}
\begin{equation}
    \mathbb{Y}_{i,j} = \mathbb{A}_{i,k,l}\mathbb{X}_{k,j}^1\mathbb{X}_{l,j}^2
\end{equation}
\begin{equation}
    \mathbb{Y}_{i,j,l} = \mathbb{A}_{i,j,k}\mathbb{X}_{k,l}^1
\end{equation}
\begin{equation}
    \mathbb{Y}_{i,k} = \mathbb{A}_{i,k}\mathbb{X}_{i,j}^1\mathbb{X}_{j,k}^2
\end{equation}
\begin{equation}
    \mathbb{Y}_{i,k} = \mathbb{A}_{i,j}\mathbb{X}_{j,k}^1
\end{equation}
\end{subequations}

We use message-passing to define sparse-dense hybrid algebra in DF as Eq.~\ref{db-view}.
\begin{equation}\label{db-view}
    Q(dst)=\oplus_{src\in Q_0(dst)}\{src, \otimes(Q_1(src,dst), Q_2(dst))\}
\end{equation}
$Q,Q_0,Q_1,Q_2$ are queries for the relevant database. We follow the idea of logical-physical storage seperation~\cite{codd1970relational}. The value of $Q(k)$ is defined as $Q(dst)=D(f(dst))$. $D$ is the relevant database of $Q$, storing $(id, value)$ in ascending order of $id$, where $id\in \mathbb{Z}$ and $value\in \mathbb{R}^n$. $dst$ is any hashable key and f is a function $K\rightarrow \mathbb{Z}$. $\oplus$ can be any commutative operation and $\otimes$ can be any function that takes two objects as input and output one object that can be operated by $\oplus$. The result of $\oplus$ is written to $f(dst)$ in $Q$. Sparse-dense hybrid algebra is sparse because $Q_0(dst)$ for all $dst$ are diverse. In other words, $Q_0(i) \bigcap Q_0(i+1) \sim \O$. Such algebra is dense because values in $D,D_1,D_2$ are scalar, dense vectors, or dense matrices. 

\begin{figure}[h]%
\centering
\includegraphics[width=0.99\textwidth]{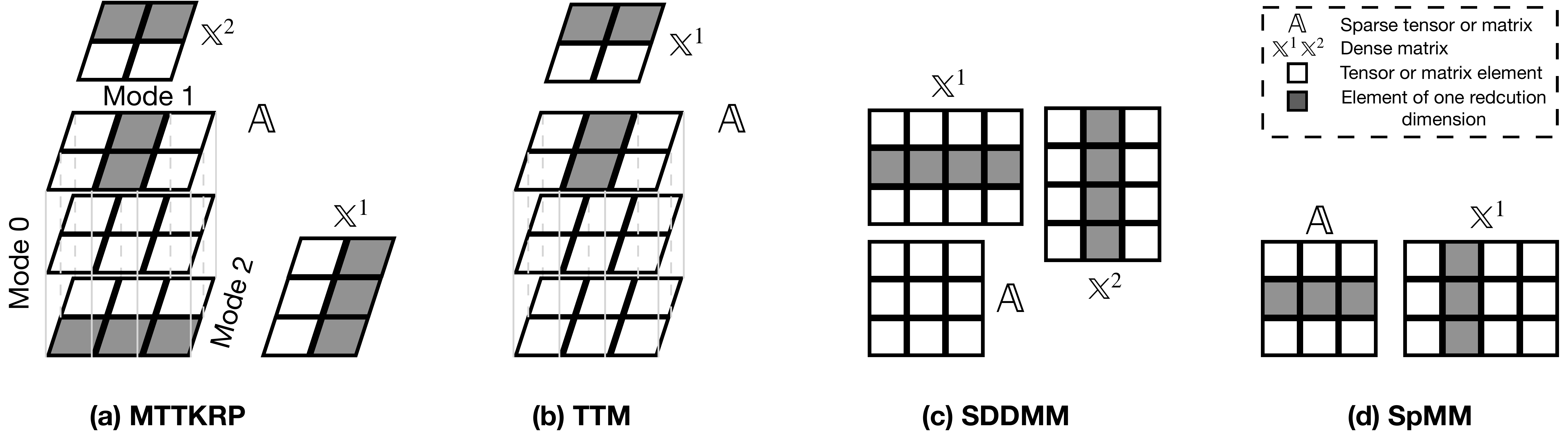}
\caption{Examples of sparse-dense hybrid algebra. The consecutive grey parallelograms or squares represent the reduction modes.}\label{fig-kernels}
\end{figure}
\begin{figure}[h]%
\centering
\includegraphics[width=0.99\textwidth]{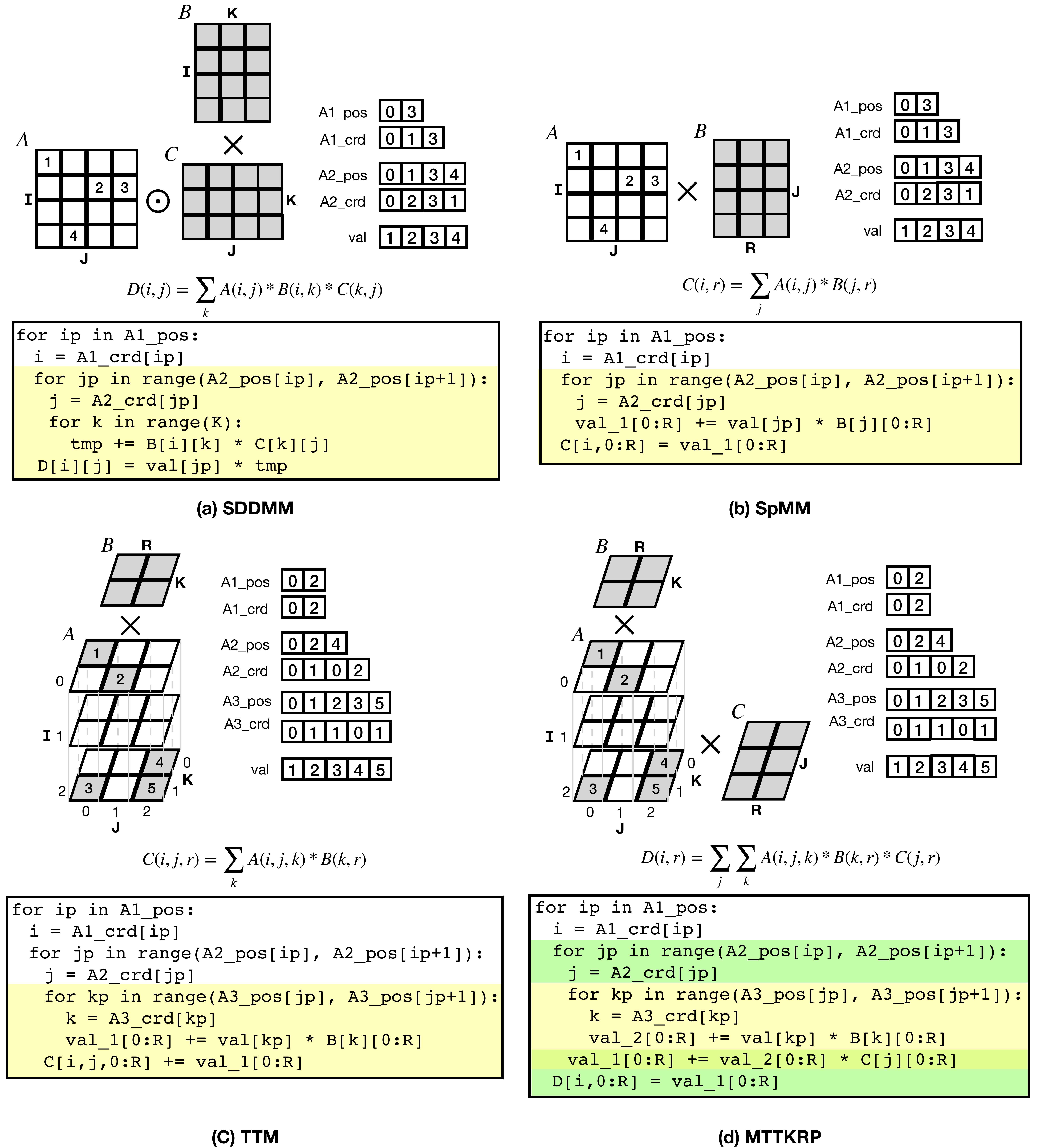}
\caption{Code examples of reduction in sparse-dense hybrid algebra in TF. The colored lines are reduction codes. MTTKRP has two levels of reduction, colored green and yellow, respectively. The overlapped region means that the first-level reduction's output serves as the second-level reduction's input. We follow the naming rules in~\cite{kjolstad:2020:phd-thesis} for the storage of $A$.}\label{fig-code}
\end{figure}
\begin{figure}[h]%
\centering
\includegraphics[width=0.99\textwidth]{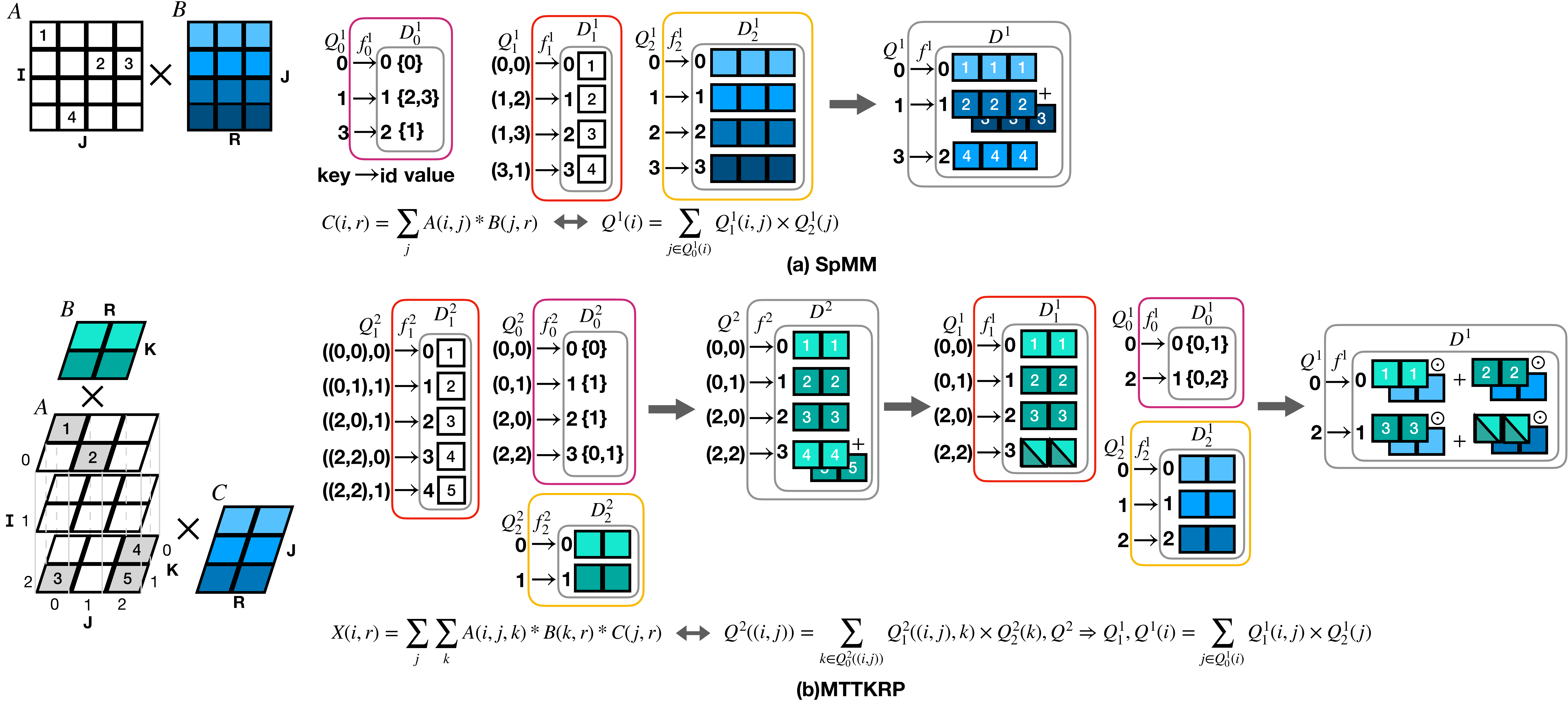}
\caption{Illustration of common reduction in MTTKRP and SpMM. The equivalent expressions of the same kernel in TF and DF are below each sub-figure.}\label{fig-redb}
\end{figure}

The core operation of sparse-dense hybrid algebra is \textit{reduction} and reduction in different kernels behaves similarly. This key observation motivates {\framework} because we only need to optimize the common reduction operations and use the compiler to optimize different sparse-dense hybrid algebra kernels automatically. For example, in TF kernels do reduction on $l,k$ dimensions in MTTKRP, $k$ in TTM, $j$ in SDDMM and SpMM. The reduction can be along one sparse and one dense dimension, as in MTTKRP, TTM, and SpMM. It can also be along two dense dimensions, as in SDDMM. Fig.~\ref{fig-kernels} illustrates these examples and highlights the reduction dimensions. We also give concrete code examples in Fig.~\ref{fig-code}. It shows that some of these kernels share common reduction codes. For example, MTTKRP contains two reductions, each behaving the same as the reduction in SpMM. 
Such property can also be illustrated in DF. As shown in Fig.~\ref{fig-redb}, for the first reduction, the value of $D_1$ both are scalar; the value of $D_2$ both are vectors. For the second reduction of MTTKRP, though the value of $D_1$ is a vector, which is different from SpMM's first reduction, $\oplus$ behaves the same because $\otimes$ here is element-wise vector product.   
\subsection{SpMM Optimization}
As explained above, the reduction is the core operation of sparse-dense tensor algebra and some kernels share the same type of reduction. Without loss of generality, we take SpMM as an example to optimize the reduction in this paper. The optimization techniques can be easily generalized to expedite other sparse-dense hybrid algebra kernels. Yang et al.~\cite{yang2018design} selects between two algorithms to achieve respectively even distribution of nnz among parallel processors and row-splitting among threads. Adaptive Sparse Tiling (ASpT)~\cite{hong2019adaptive} aims at improving data locality and thus reduces the total number of accesses to global memory. Ge-SpMM~\cite{huang2020ge} proposes Coalesced Row Caching (CRC) method to enable coalesced memory access to both sparse and dense matrices and Coarse-grained Warp Merging (CWM) method for SpMM merging workloads from different warps to reuses loaded sparse matrix. Mehrabi et al.~\cite{mehrabi2021learning} proposes several row permutation strategies for CSR format to enhance load balance and data locality. DA-SpMM~\cite{dai2022heuristic} is a data-aware kernel selector among 8 algorithms according to 3 dimensions in the space dealing with dynamic input data.
\subsection{Sparse Compilers}
The complexity of optimizing sparse tensor algebra comes from four directions: data, data format, algebra, and hardware. Researchers often develop a technique for one data format, one algebra, and one hardware. Such a library method heavily relies on experts and engineering work~\cite{guennebaud2010eigen,naumov2010cusparse,wang2014mkl}. However, sparse compilers can extremely reduce such engineering burden and boost innovation in this area. Unlike the library method, sparse compilers aim to use \textbf{one} monolithic theory to express all data formats, all algebras, and provide flexible user interface, which enables users to explore the optimization space given data and hardware.
Research on sparse compilers can be divided into two categories: (1) \textit{Pass-oriented}. Given the imperative code, design compilation passes to optimize the code~\cite{bik1993compilation,venkat2015loop,strout2018sparse}. (2) \textit{Language-oriented}. View sparse compiler as a programming language and design lowering and scheduling process~\cite{ye2022sparsetir,bik2022compiler,kjolstad:2020:phd-thesis}. Especially, TACO is a fundamental breakthrough on this problem. To the best of our knowledge, it is the first to propose a practical sparse compilation theory. MLIR sparse dialect~\cite{bik2022compiler} implements TACO's sparse compilation theory as MLIR dialect. SparseTIR~\cite{ye2022sparsetir} follows the design philosophy of TensorIR~\cite{feng2022tensorir}, but it still uses some of the TACO's concepts such as position and coordinate space. TACO also motivates innovations on accelerators for sparse tensor algebra~\cite{qin2022HardTACO}.
\subsection{TACO}
\begin{figure}[h]%
\centering
\includegraphics[width=0.9\textwidth]{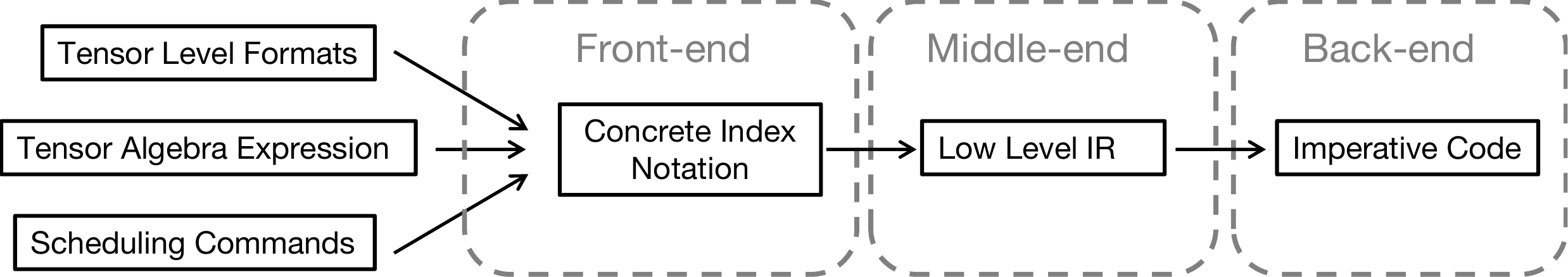}
\caption{Overview of the TACO workflow}\label{fig0}
\end{figure}

TACO (The Tensor Algebra Compiler) is a fast and versatile compiler-based library for sparse linear and tensor algebra~\cite{kjolstad:2017:taco,kjolstad:2019:workspaces,kjolstad:2020:phd-thesis,senanayake:2020:scheduling}. TACO has three types of inputs: a tensor algebra expression (in an Einstein summation notation or reduction notation); level formats of input and output tensor; schedule commands. We will introduce TACO in the front-end, middle-end, and back-end order. The workflow of TACO is illustrated in Fig.~\ref{fig0}.

\subsubsection{Front-end}\label{rewriter}
At the front-end, the tensor algebra expression is concretized to concrete index notation~\cite{kjolstad:2019:workspaces}. The concrete index notation (CIN) is a language that describes the execution of a tensor algebra. Unlike bare tensor algebra expression, CIN describes the loop, index variables relations, workspace, hardware platform, etc. Schedule commands transform the CIN. For example, a \textit{precompute} schedule will add a $where$ statement to the CIN. Though TACO provides a clean and powerful scheduling API to transform CIN, the user can still change the CIN directly. TACO provides a match function that can take lambda expression as input. The function can modify CIN when it meets a specific type of CIN node or a pattern of CIN nodes. Moreover, users can define a child class of IndexNotationRewriter that can directly rewrite the CIN. Such technique is used to implement {\newabs}.
\subsubsection{Middle-end}
At the middle-end, CIN will be transformed to imperative IR (or low level IR (LLIR)). LLIR describes the basic blocks, for example, for-loop, while-loop, and if-statement. LLIR is almost the executable code. The output of the middle-end is a chain of LLIR. The sparse iteration theory~\cite{kjolstad:2020:phd-thesis} guides the CIN to LLIR process. It ensures that different tensors only coiterate over elements that can generate non-zero output. Specifically, TACO designs lower functions for every statement of CIN and lattices in the sparse iteration space. However, current lower functions only assume serial reduction is done on the compressed level of sparse tensors. We will break the serial code assumption to implement {\newabs}. Moreover, we suggest that more flexible or even user-defined lowerers should be designed in the future.
\subsubsection{Back-end}
At the back-end, LLIR will be transformed to code for different backends. In this paper, we target the CUDA code generation. TACO CUDA code generator has some assumptions that previous papers did not thoroughly explore. TACO deals with CUDA code generation in a nested loop favor~\cite{senanayake:2020:scheduling}. Moreover, it only generates one dimension of block and thread. That is, it only has blockIdx.x and threadId.x. When the index variable of a for-loop LLIR is bound on the GPUBlock, it will use blockIdx.x to index this index variable. In the CPU case, it will emit a real for-loop. Such variable is assumed to increment by 1. Index variables bound on GPUWarp and GPUThread are assumed to be the outer and inner variables of threadIdx.x. The tile size depends on the index variable on GPUThread. The mixture of tiling and synchronization semantics of GPUWarp loses some optimization opportunities. We will discuss this later and improve it in our implementation.

\section{\Framework}\label{sec:framework}
\subsection{Computation unit model}
We observe that the core operation of sparse-dense hybrid algebra is the reduction. Therefore, the core of our model is \textit{how many data are reduced and are synchronized in which way}. We model the atomic computation unit as \textbf{thread}. A thread executes a serial program. All threads execute the same program independently with each own's input data and are distinguished by threadId. Threads can do synchronization in groups with \textit{reduction parallelism} of 2,4,8,16, or 32. We model GPU computation as \textit{unlimited} parallel threads and define the number of threads as \textit{resource parallelism} that GPU can provide. We do not consider the shared memory, grid level, and the mapping of the thread block or the streaming processor. Instead, we view them as reasonable implementation details after the basic parallel pattern is decided. In other words, there can be many kinds of implementation for each algorithm in {\framework}. In this sense, {\framework} can encourage more GPU optimization innovation.
\subsection{Overview of {\framework}}
 To define the parallel pattern concretely, we propose \textit{atomic parallelism}. A program with \textit{atomic parallelism} cannot be paralleled anymore. In other words, a thread at least executes the amount of data denoted by atomic parallelism. Formally, atomic parallelism is defined as the Cartesian product of \textit{minimal data}. Minimal data is the minor data of one category a thread can execute. Atomic parallelism can be used to construct the optimization space of any sparse-dense hybrid algebra under the GPU model, but we focus on SpMM in this paper. 
 
 Indeed, tiling, manipulating shared memory, and thread mapping~\cite{hidayetouglu2020scale,mehrabi2021learning,xin2021fast,huang2020ge} are also important for SpMM on GPU. They are crucial for SpMM, especially with many dense columns(usually more than 128 columns), because the computation will be more \textit{workload} intensive and bounded by the memory access for dense columns. However, we focus on SpMM with fewer dense columns(usually less than 8 columns), which are more \textit{balance} intensive and bounded by the maximum warp execution cycles. 
 
 SpMM has two orthogonal atomic parallelisms: minimal data can be (1) $\{\frac{1}{g},1,g\}$ non-zeros of the sparse matrix and $\{\frac{1}{c},1,c\}$ columns of the dense matrix; (2) $\{\frac{1}{g},1,g\}$ rows of the sparse matrix and $\{\frac{1}{c},1,c\}$ columns of the dense matrix. $c\in \mathbb{Z^+}$ and $g\in \mathbb{Z^+}$ are tunable parameters. Though they can be 1, they have different meanings from 1, because they are \textit{tunable}. Therefore, the atomic parallelism space of SpMM is described in $<x\,nnz , y\,col>$ or $<x\,row, y\,col>$. Resource parallelism only multiplies one element of the atomic parallelism. For example, given resource parallelism $r$, the amount of executed data equals  $<r\times x\,nnz , y\,col>$ or $<x\,nnz , r\times y\,col>$. Besides, a fractional amount of data means multiple threads may execute on the same datum. For example, $<\frac{1}{g}\,row, 1\,col>$ means that $g$ threads execute the same row collaboratively.
 
\subsection{SpMM optimization space formalization}
\begin{figure}[h]%
\centering
\includegraphics[width=0.9\textwidth]{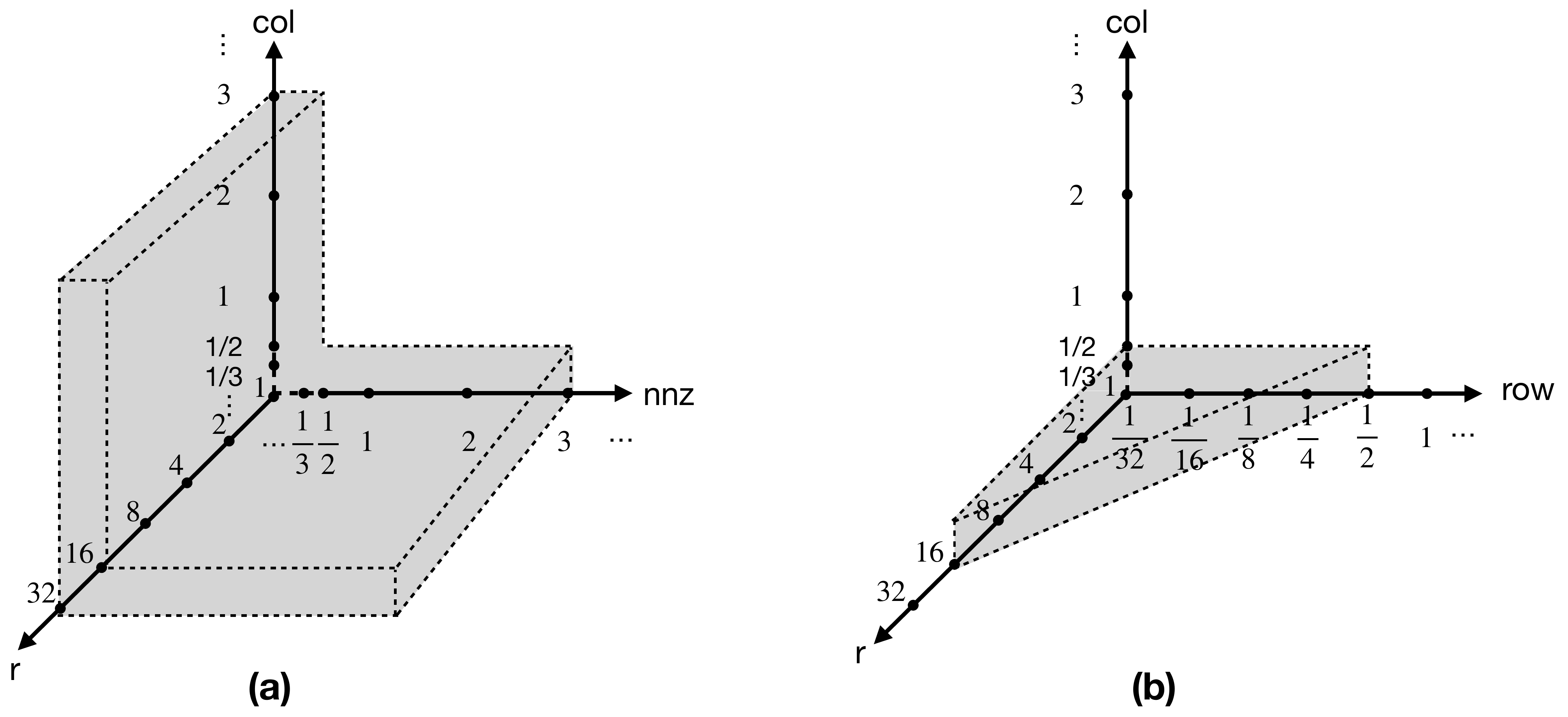}
\caption{SpMM optimization space. The grey area is illegal. The dashed line part of the axis represents hardware dependent end of the axis.}\label{fig-space3d}
\end{figure}

\begin{figure}[h]%
\centering
\includegraphics[width=0.99\textwidth]{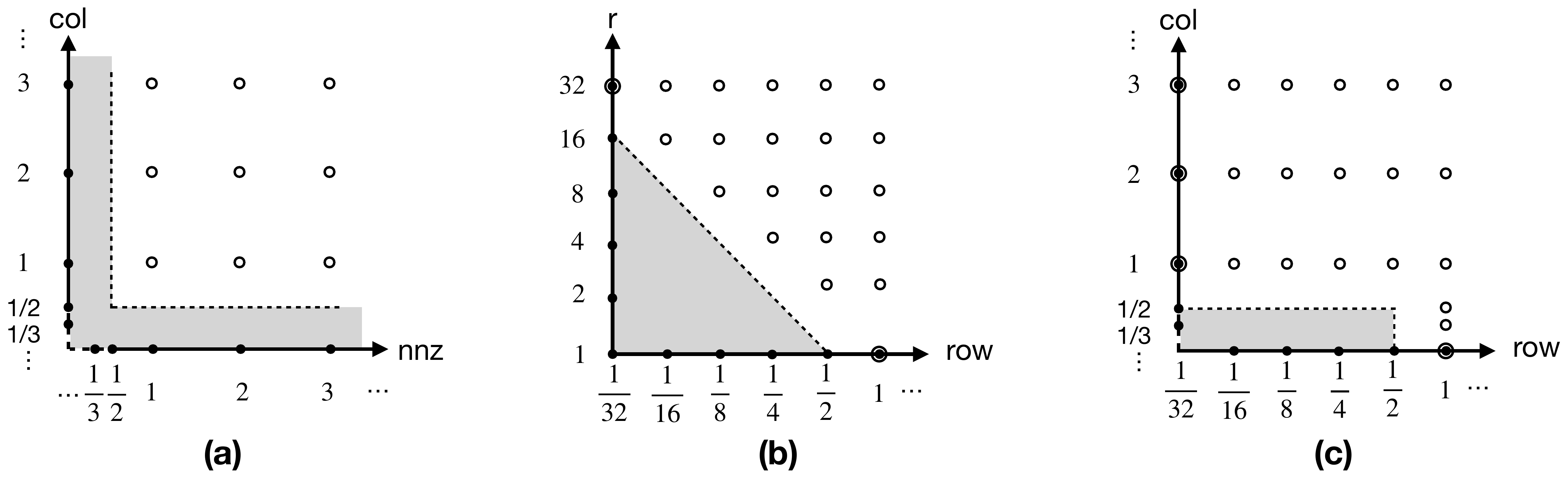}
\caption{Projections of SpMM optimization space. Grey areas are illegal and hollow circles are legal points. Sub-figures (a), (b), and (c) correspond to Rule 1, 2, and 3 respectively.}\label{fig-space}
\end{figure}

We use atomic parallelism and reduction parallelism $\{<...>,r\}$ to define an SpMM kernel. $<...>\in \{\frac{1}{g},1,g\}nnz \times \{\frac{1}{c},1,c\} col $ or $\{\frac{1}{g},1,g\}row \times \{\frac{1}{c},1,c\} col$. They describe the minimal data. And the \textit{reduction parallelism} $r\in\{2,4,8,16,32\}$ assigns how many threads are synchronized each time. Fig.~\ref{fig-space3d} illustrates the SpMM optimization space. 
However, not all points in the atomic parallelism space are legal in optimization space. Fig.~\ref{fig-space} illustrates the details of space pruning. There are three rules for legal points:
\begin{enumerate}[(1)]
\item$\{<\frac{1}{g}\,nnz , x\,col>,r\}$, $\{<x\,nnz , \frac{1}{c}\,col>,r\}$ are illegal because one non-zero must by multiplied by at least one element in the dense matrix. 
\item$\{<\frac{1}{g}\,row, x\,col>,r\}(\frac{r}{g}<1)$ is illegal because parallel reduction only has one writeback thread.
\item$\{<\frac{1}{g}\,row , \frac{1}{c}\,col>,r\}$ is illegal because it conflicts with the rule that resource parallelism only multiplies one element of the atomic parallelism.
\end{enumerate}

The state-of-the-art algorithm space, DA-SpMM~\cite{dai2022heuristic} is in the {\framework} design space. It proposes a three-dimensional SpMM algorithm design space. We claim that the design space of DA-SpMM is included in the atomic parallelism space. To be specific, EB+PR is $\{<1\,nnz , c\,col>,32\}$, RB+PR is $\{<\frac{1}{32}\,row, c\,col>,32\}$, EB+SR is $\{<32\,nnz,c\,col >,1\}$, and RB+SR is $\{<1\,row,c\,col >,1\}$. $c$ means coarsen factor, $g$ means group size. Though real CUDA code with $1\,row$ or $1\,nnz$ may have minimal data greater than one because of limited resource parallelism, we still label the algorithm as $1\,row$ or $1\,nnz$. The RM/CM is the implementation detail and is included in {\framework} in theory.

\section{\Newabs}\label{sec:newabs}
\subsection{Current warp-level abstraction}
\begin{figure}[h]%
\centering
\includegraphics[width=0.99\textwidth]{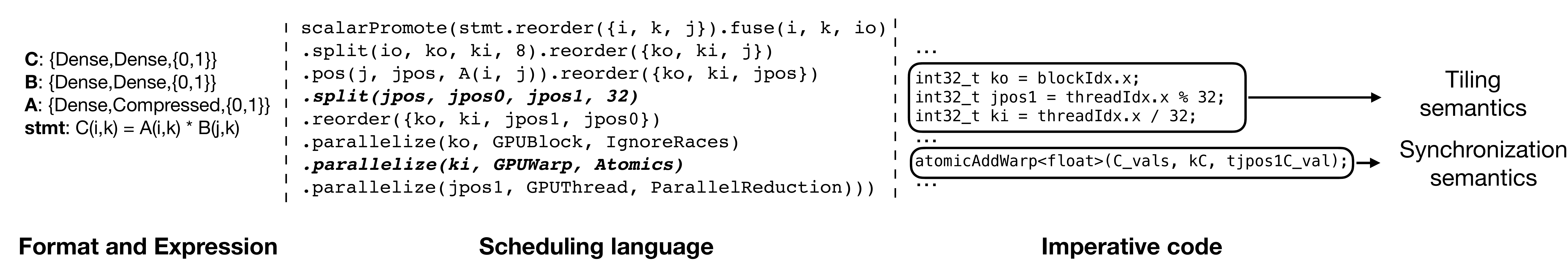}
\caption{Tiling and synchronization semantics of GPU Warp in TACO}\label{fig-sem}
\end{figure}

Current sparse tensor compilers with CUDA backend take warp as the rank of a thread (\textit{tiling}), a particular parallel unit (\textit{synchronization}) or just a hardware instruction. For example, TACO assumes warp and thread to be the outer and inner loop, and the $warpSize$ depends on the split factor. It should be noted that no synchronization behavior is assumed in this case. TACO also takes the 32-thread warp reduction as atomic addition at the GPUWarp parallel unit and assumes users will split the last level loop with $warpSize=32$. In this case, CUDA warp is taken as a for-loop with extent $warpSize$ and incremental step $1$. Then they will emit CUDA warp primitives such as \textit{\_\_shfl\_down\_sync} to do the reduction. Fig.~\ref{fig-sem} illustrates TACO's current GPU Warp semantics. On the contrary, TVM\cite{chen2018tvm} only binds on thread and block level and does not assign any synchronization on the warp level. Instead, it takes 32 as a hardware feature and uses such intrinsic to fill in schedule parameters in auto-scheduler. Besides, it also uses warp as a memory load unit in TIR\cite{chen2018tvm}. 
\subsection{Overview of {\newabs}}
However, at least two existing assumptions should be improved for sparse compilers. First, the \textit{tiling} and \textit{synchronization} semantics of warp should be explicitly separated. As shown in {\framework}, the atomic and reduction parallelism can be different, and reduction parallelism is not necessarily 32. Second, synchronization semantics should be able to express various reduction strategies and flexible reduction granularity, instead of just parallel reduction for 32 threads. As shown in {\framework}, $\{<1\,nnz , c\,col>,n\}$ requires synchronization of $n$ threads with row number of their own. Therefore, the warp reduction should be able to reduce to different outputs instead of only one. Such change not only calls for changing the hand-coded warp level reduction functions but also for elevating the reduction pattern to higher-level compiler passes. Such semantics lifting calls for a new organization of basic blocks, new control flow, and new user-level APIs.

\subsection{Relationship between {\newabs} and {\framework}}
{\Framework} models the optimization space of sparse-dense hybrid algebra from the reduction view. We use this model to propose new optimization techniques. As shown in Section~\ref{sec:bg}, reduction is the key operation of sparse-dense hybrid algebra, which contains many different tensor algebras such as SpMM, SDDMM, MTTKRP, and TTM. Based on this observation, we define and explain {\newabs} in Section~\ref{sec:framework}, using SpMM as an example. We show that~\ref{sec:framework} opens new optimization space for SpMM. Such benefit can be generalized to other sparse-dense hybrid algebra. However, it requires repetitive engineering efforts to optimize case by case. In response to this issue, we propose {\newabs}, a new abstraction for sparse compilers to ship performance benefits brought by {\framework} to users with only several lines of code changed on the user side. 

\begin{figure}[h]%
\centering
\includegraphics[width=0.9\textwidth]{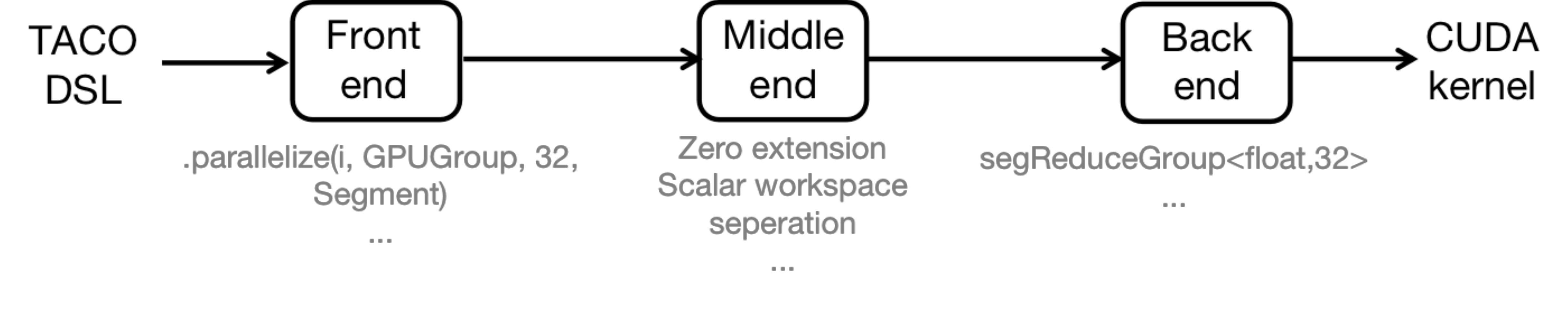}
\caption{Overview of {\newabs} in the TACO workflow}\label{fig1}
\end{figure}

In summary, we propose that sparse compilers for GPU should have abstraction {\newabs}, that is, a \textit{warp} that takes the \textit{tiling} semantics, and a \textit{group} that does different types of reduction \textit{synchronization}.  We will use TACO\footnote{We build on commit d0654a8 \url{https://github.com/zhang677/taco/tree/d0654a84137169883973c40a951dfdb89883fd9c}} to illustrate how to implement {\newabs}, but other sparse compilers can also integrate {\newabs}. Fig.~\ref{fig1} illustrates the workflow.
\section{{\Newabs} for TACO}\label{sec:newabs4taco}

The original \textit{parallelize} transformation is defined as \textit{parallelize(IndexVar i, ParallelUnit pu, OutputRaceStrategy rs)}~\cite{senanayake:2020:scheduling}. The transformation does parallel execution on IndexVar $i$, using ParallelUnit $pu$. And OutputRaceStrategy $rs$ describes the data races during reductions. For GPU, $pu$ can be GPUThread, GPUWarp, and GPUBlock. $rs$ can be NoRaces, IgnoreRaces, and Atomics. We propose two new designs to TACO:

\begin{enumerate}[1.]
\item We add a new PrallelUnit, \textit{GPUGroup}, to  the \textit{parallelize} transformation, and change the semantics of ParallelUnit \textit{GPUWarp}. 
\item We break the assumption that other transformations other than parallelize assumes serial code and design a new lower process to enable segment reduction.
\end{enumerate}
\subsection{New parallelize transformation}
We assign the \textit{tiling} semantics to GPUWarp and its \textit{Atomic OutputRaceStrategy} will only serve to direct the lowering function instead of synchronization semantics. Because GPUWarp now only serves as the outer loop of tiling on threadIdx, it does not have \textit{Atomic} semantics. Meanwhile, we add \textit{GPUGroup} which has \textit{ReductionStrategy} and \textit{GroupSize} attributes instead of \textit{OutputRaceStrategy}. \textit{ReductionStrategy} describes the group's reduction type, and \textit{GroupSize} assigns the reduction parallelism.
\subsection{Reduction semantics elevation}
TACO assumes that a sparse algebra compiler should do it best to ensure that only elements that can generate non-zero output will be calculated~\cite{kjolstad:2020:phd-thesis}. However, we point out that this assumption is not necessarily valid. The previous assumption is the best option for performance because the sparse iteration space theory is built on the assumption that the code runs serially. For CUDA code, however, such assumption is broken, which we term as \textit{zero extension}. Zero extension means that some ``out-of-bound'' reduction can be allowed in the sparse iteration theory because it can later be executed by some warp primitives faster than for-loop. 
\subsection{Segment reduction lowering}
\begin{minipage}[t]{0.47\linewidth}
\begin{lstlisting}[language=c++,caption={Original CUDA code },captionpos=b,label={lst:old-code},basicstyle=\footnotesize]
//Original CUDA code 
for(k=0;k<B2_dimension;k++){
  pA2_begin=i_blockStarts[block];
  pA2_end=i_blockStarts[block+1];
  fposA=block*256;
  i_pos=taco_binarySearchBefore(
  A2_pos,pA2_begin,pA2_end,fposA);
  i=i_pos;
  fposA=block*256+fpos1;
  if(fposA>=A2_pos[A1_dimension])
    break;
  f=A2_crd[fposA];
  kB=f*B2_dimension+k;
  while(fposA==A2_pos[i_pos+1]){
    i_pos=i_pos+1;
    i=i_pos;
  }
  kC=i*C2_dimension+k;
  float val=0.0;
  val=A_vals[fposA]*B_vals[kB];
  atomicAdd(&C_vals[kC],val);
}
\end{lstlisting}
\end{minipage}%
\hfill\vrule\hfill
\begin{minipage}[t]{0.47\linewidth}
\begin{lstlisting}[language=c++,caption={Modified CUDA code },captionpos=b,label={lst:new-code},basicstyle=\footnotesize]
//Modified CUDA code
for(k=0;k<B2_dimension;k++){
  pA2_begin=i_blockStarts[block];
  pA2_end=i_blockStarts[block+1];
  fposA=block*256+fpos1;
  i_pos=taco_binarySearchBefore(
  A2_pos,pA2_begin,pA2_end,fposA);
  i=i_pos;
  float val=0.0;
  if(fposA>=A2_pos[A1_dimension])
    val=0;
  else{
    f=A2_crd[fposA];
    kB=f*B2_dimension+k;
    while(fposA==A2_pos[i_pos+1]){
      i_pos=i_pos+1;
      i=i_pos;
    }
    val=A_vals[fposA]*B_vals[kB];
  }
  kC=i*C2_dimension+k;
  segReduceWarp<float,32>(C_vals,
  kC,val);
}
\end{lstlisting}
\end{minipage}%

Listing~\ref{lst:old-code} and Listing~\ref{lst:new-code} show the difference between codes generated by the original TACO and the modified TACO. They use the same schedule, except that code on the right uses segment reduction of GPUGroup with size 32.

\textbf{scalar workspace}. TACO assumes that the statement and the assignment of \textit{scalar workspace}~\cite{kjolstad:2019:workspaces} are in the same basic block. However, this assumption is so strong that it restricts the expressive power of TACO. For example, in $\{<1\,nnz , c\,col>,32\}$ the scalar workspace should be assigned in a basic block belonging to an \textit{else} but stated in the same context with reduction of scalar workspace, outside the assignment basic block. 

\textbf{Macro instruction}. It is important to emit code in a modular way. Therefore, we design two new \textit{macro instructions} \textit{atomicAddGroup$<$T,G$>$(T* array, int idx, T value)} and \textit{segReducWarp$<$T,G$>$(T* array, int idx, T value)}. They are template device functions that takes in the output array, the index of the output and the value reduced to the output\footnote{We do not actually integrate these macro instructions into TACO, because it is fairly straightforward and purely engineering. When testing the kernels, we just replace the atomicAdd with the new macro instructions. We open-source the modified TACO \url{https://github.com/zhang677/taco/tree/parallelreduction}.}. They will do some kind of reduction on $G$ threads, and $G$ equals $GroupSize$. They will be stated in the header file and used as macro instructions in the final CUDA code. In fact, we borrow the $group$ concept from the \textit{cooperative group} in CUDA. Since CUDA 11.0, it has supported an easy-to-use API called cooperative group \footnote{\url{https://docs.nvidia.com/cuda/cuda-c-programming-guide/index.html\#cooperative-groups}} that makes it only one-line-code effort to change reduction granularity to less than 32 threads.

\section{TACO's support for four SpMM algorithms}\label{sec:tacospmm}
This section will illustrate the {\framework} design space and our implementation of {\newabs}. We first reexamine two SpMM algorithms proposed by TACO~\cite{senanayake:2020:scheduling}. They use TACO to generate $\{<g\,nnz, c\,col>,1\}$ and $\{<x\,row,c\,col >,1\}$. We then use another two examples, $\{<\frac{1}{g}\,row, c\,col>,r\}$ and $\{<1\,nnz , c\,col>,r\}$ to illustrate how the CIN is changed. The tensor algebra expression is $C(i,k) = A(i,j) * B(j,k)$. $A$'s first level is dense and the second level is compressed. $B$ and $C$ are both dense matrices. $A,B$, and $C$ all are row-major. We assume $N=4$ and that thread per block (resource parallelism $p$) equals 256. We explicitly fill $p,g,N,c$ into the CIN to show their arithmetic relations with CIN parameters. The actual CIN will not have undetermined variables.
\subsection{TACO SpMM reexamination}
Currently, TACO supports two algorithms in {\framework}. They don't need synchronization semantics and only tune on the tiling semantics. The implementation by TACO is shown in Listing ~\ref{old_1} and ~\ref{old_2}. They force the synchronization granularity to be 1 which presents limited capability in reduction.

Concrete Index Notation for $\{<g\,nnz, c\,col>,1\}$ is :

\begin{minipage}{\hsize}%
\lstset{frame=single,framexleftmargin=-1pt,framexrightmargin=-17pt,framesep=12pt,linewidth=0.97\textwidth,language=pascal,caption={CIN for $\{<g\,nnz, c\,col>,1\}$},captionpos=b,basicstyle=\footnotesize}
\begin{lstlisting}[label={old_1}]
suchthat(forall(block,forall(warp,forall(thread,
forall(dense_val,where(C(i,k)+=tnnzC,forall(nnz, 
tnnzC+=A(i,j)*B(j,k)))),GPUThread,Atomics), 
GPUWarp,NoRaces),GPUBlock,NoRaces), 
fuse(i,j,f) and pos(f,fpos,A(i,j)) and 
split(fpos,block,fpos1,(p*g/(N/c))) and 
split(fpos1,warp,nnz,g) and split(k,ko,thread,c)
and bound(ko,dense_val,N/c,MaxExact))
\end{lstlisting}
\end{minipage}
Actually, TACO's \textit{precompute} schedule fails to generate this CIN, so we use the IndexNotationRewriter technique mentioned in section \ref{rewriter} to get the CIN above. In the evaluation section of~\cite{senanayake:2020:scheduling} it assumes $N=128,g=16,c=4,p=512$, which is a point in the $\{<g\,nnz, c\,col>,1\}$. 

Concrete Index Notation for $\{<g\,row,c\,col >,1\}$ is :

\begin{minipage}{\hsize}%
\lstset{frame=single,framexleftmargin=-1pt,framexrightmargin=-17pt,framesep=12pt,linewidth=0.97\textwidth,language=pascal,caption={CIN for $\{<g\,row,c\,col >,1\}$},captionpos=b,basicstyle=\footnotesize}
\begin{lstlisting}[label={old_2}]
suchthat(forall(block,forall(warp,forall(row,
forall(thread,forall(col,where(C(i,k)+=tjC,
forall(j,tjC+=A(i,j)*B(j,k)))),GPUThread,NoRaces)),
GPUWarp,NoRaces),GPUBlock,NoRaces),split(i,block,io,
p*g/(N/c))and split(io,warp,row,g) and split(k,ko,col,c)
and bound(ko,thread,N/c,MaxExact))
\end{lstlisting}
\end{minipage}
The generated code can be directly executed. In the evaluation section of~\cite{senanayake:2020:scheduling} it assumes $N=128,g=1,c=4,p=512$, which is also a point in the $\{<g\,nnz, c\,col>,1\}$. These two algorithms only use the \textit{tiling} semantics of GPUWarp.

\subsection{Two new algorithms}
We introduce two algorithms to overcome the restricted scheme forced by TACO to improve workload balance.
The algorithms provide functionality to change group size and reduction strategy through tuning nnz and rows.
Listing~\ref{algm_1} and~\ref{algm_2} show the implementation.


Concrete Index Notation for $\{<\frac{1}{g}\,row, c\,col>,r\}$ is :

\begin{minipage}{\hsize}%
\lstset{frame=single,framexleftmargin=-1pt,framexrightmargin=-17pt,framesep=12pt,linewidth=0.97\textwidth,language=pascal,caption={CIN for $\{<\frac{1}{g}\,row,c\,col>,r\}$},captionpos=b,basicstyle=\footnotesize}
\begin{lstlisting}[label={algm_1}]
suchthat(forall(ko,forall(warp,forall(kii,where(C(i,k)+=tjpos1C,
forall(jpos1, forall(jpos0,tjpos1C+=A(i,j)*B(j,k)),GPUThread,
ParallelReduction))),GPUWarp, Atomics),GPUBlock,NoRaces),
fuse(i,k,io) and split(io,ko,ki,c*p/g) and split(ki, warp,kii,c) 
and pos(j,jpos,A(i,j)) and split(jpos,jpos0,jpos1,g) and
parallelize(jpos1,GPUGroup,r,Atomics))
\end{lstlisting}
\end{minipage}
We find that TACO can support $g=32, r=32$, but it is not explored in the autoscheduling paper\footnote{\cite{senanayake:2020:scheduling}'s authors shared their \href{https://drive.google.com/file/d/1qZbP7tY5N35N54JlmYkBHxY97HbgFSHE/view?usp=sharing}{code} with us. We also use a similar code base to test our kernels in Section \ref{sec:exp}}. GPUGroup is bound on the indexVar that does the reduction. Generated macro-instruction, \textit{atomicAddWarp$<$Type$>$}, is changed to \textit{atomicAddGroup$<$Type, G$>$} to enable more fine-grained thread synchronization.

Concrete Index Notation for $\{<1\,nnz , c\,col>,r\}$ is :

\begin{minipage}{\hsize}%
\lstset{frame=single,framexleftmargin=-1pt,framexrightmargin=-17pt,framesep=12pt,linewidth=0.97\textwidth,language=pascal,caption={CIN for $\{<1\,nnz , c\,col>,r\}$},captionpos=b,basicstyle=\footnotesize}
\begin{lstlisting}[label={algm_2}]
suchthat(forall(block,forall(warp,forall(ki,forall(fpos1,where(
C(i,k)+=tmp,tmp=A(i,j)*B(j,k)),GPUThread,Atomics)),GPUWarp,NoRaces),
GPUBlock,IgnoreRaces),fuse(i,j,f) and pos(f,fpos,A(i,j)) and
split(fpos,block,fpos1,p/(N/c)) and split (k,ko,ki,c) and bound(ko,
warp,N/c,MaxExact) and parallelize(jpos1,GPUGroup,r,Segment))
\end{lstlisting}
\end{minipage}
This algorithm has no counterpart in the original TACO. We change the originally emitted \textit{atomicAdd} to \textit{segReduceGroup$<$Type,G$>$}, and the grouped segment reduction is done in the macro instruction. The lowerer of scalar workspace is changed to emit the code ready for segmented reduction.
\section{Evaluation}\label{sec:exp}
\textbf{Experiment settings.} We evaluate the implementation and the generalization on three architectures:
\begin{itemize}
\item NVIDIA RTX 3090. Compute Capability 8.6 (68 Ampere SMs at 1.395 GHz, 24 GB GDDR6x, 936 GB/s bandwidth).
\item NVIDIA RTX 2080. Compute Capability 7.5 (46 Turing SMs at
1.515 GHz, 8 GB GDDR6, 448 GB/s bandwidth).
\item NVIDIA Tesla V100. Compute Capability 7.0 (80 Volta SMs at
1.370 GHz, 16 GB HBM2, 900 GB/s bandwidth). 
\end{itemize}
We use NVCC 11.6 and CUDA 11.6 with the same compilation flags as \cite{senanayake:2020:scheduling} when testing TACO and the same compilation flag as \cite{dai2022heuristic} when testing the generalized tuning. We carry 25 tests for each kernel to get the average execution time when evaluating TACO's generated CUDA kernels. We use nsight-compute\footnote{\url{https://docs.nvidia.com/nsight-compute/NsightCompute/index.html}} to get the execution time of tuned dgSPARSE kernels. We use the same sparse matrices as~\cite{dai2022heuristic}. We evaluate on three different architectures to show that our techniques are not limited to specific traits on certain generations of GPU, but are valid on common SIMT architectures.
\subsection{Performance of two new algorithms for TACO}
This experiment aims to prove that {\newabs} can improve the sparse compiler's expression ability and boost the performance of SpMM kernels generated by TACO. The dense input matrices have $N=4$\footnote{We open source the testing code at \url{https://github.com/zhang677/segTACO}.}. 

\textbf{Against the static group size 32.} We use $\{<\frac{1}{g}\,row, c\,col>,r\}$ to show the improvement brought by flexible group size $r$. Current TACO only supports $g=32,r=32$, so we keep the same $g$ with TACO and change $r$. In Table~\ref{tab-static32} we show that $r=8$ and $r=4$ can bring over 2.0x speedup on average. We also measure the \textit{normalized speedup}. Normalized speedup of $A$ over $B$ means that if $A$ performs better than $B$, we count the speedup; otherwise, we assume the user can choose the better algorithm, and the speedup is counted as 1.  
\begin{table}[h]
\begin{center}
\begin{minipage}{\textwidth}
\caption{Flexible group size speedup}\label{tab-static32}%
\begin{tabular*}{\textwidth}{@{\extracolsep{\fill}}lllll@{\extracolsep{\fill}}}
\toprule
Hardware & $r=8$  & $r=8$ norm & $r=4$  & $r=4$ norm\\
\midrule
RTX 2080    & 2.451   & 2.478 & 2.456 & 2.483 \\
RTX 3090    & 2.236   & 2.284  & 2.259 & 2.307 \\
Tesla V100    & 2.086  & 2.143  & 2.094 & 2.150 \\
\botrule
\end{tabular*}
\end{minipage}
\end{center}
\end{table}

\textbf{Against the original reduction.} We use $\{<1\,nnz , c\,col>,r\}$ to illustrate the speedup brought by flexible reduction. Because they have different data types (nnz vs. row), we control $c$ and $r$, and compare the execution of $\{<1\,nnz , c\,col>,r\}$ with the best $g$ configuration of $\{<\frac{1}{g}\,row , c\,col>,r\}$ each dataset. We only do this experiment on RTX 3090 and record the normalized speedup here. In Table~\ref{tab-red} we show that segment reduction can bring up to 1.3x speedup over atomicWarp reduction. Limited by the number of threads per warp in GPU, $r$ can only be $1,2,4,8,16,32$. Therefore, users can try these values to tune $r$ in practice.

\begin{table}[h]
\begin{center}
\begin{minipage}{\textwidth}
\caption{Segment reduction normalized speedup}\label{tab-red}%
\begin{tabular*}{\textwidth}{@{\extracolsep{\fill}}lllll@{\extracolsep{\fill}}}
\toprule
c & r=4  & r=8 & r=16 & r=32\\
\midrule
1   & 1.008   & 1.025  & 1.085 & 1.272 \\
2    & 1.019   & 1.045  & 1.102 & 1.291 \\
4   & 1.063   & 1.095  & 1.205 & 1.381 \\
\botrule
\end{tabular*}
\end{minipage}
\end{center}
\end{table}

\begin{figure}[ht]%
\centering
\includegraphics[width=0.99\textwidth]{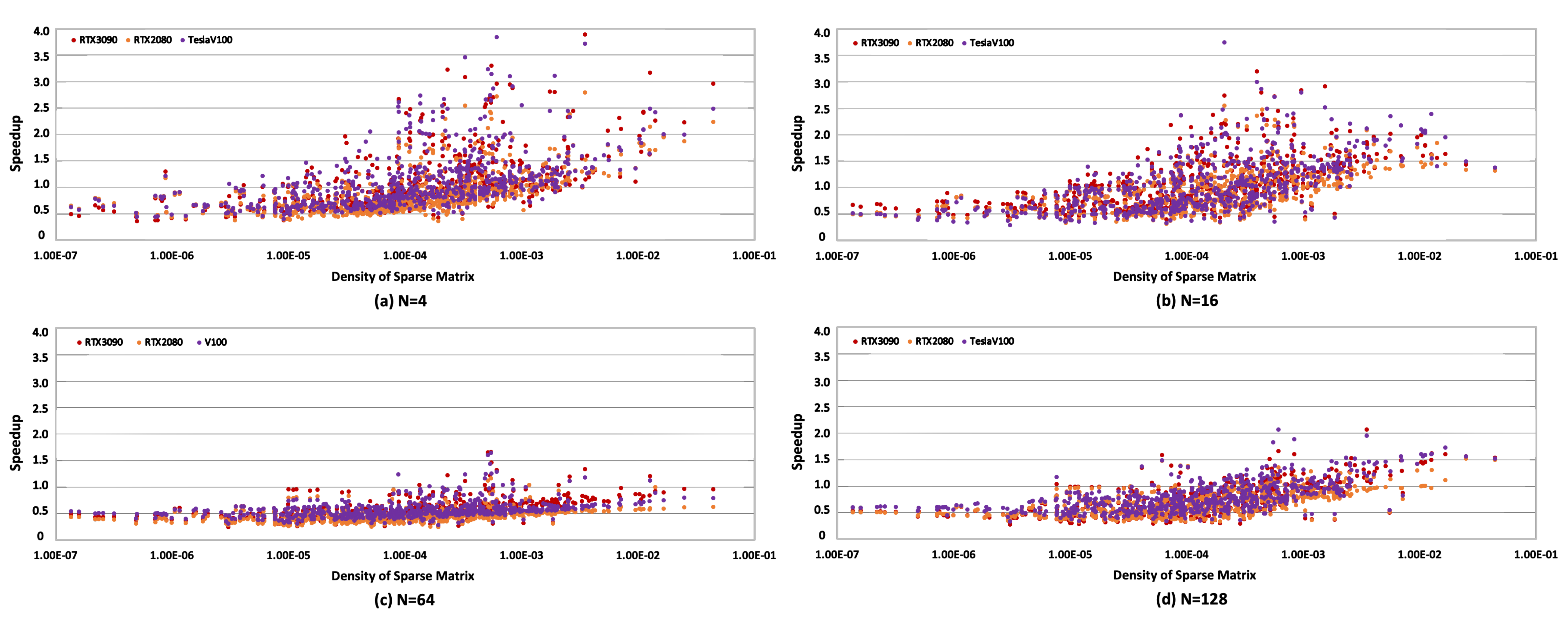}
\caption{Newly generated SpMM kernels performance compared with original TACO's best SpMM kernel for different number of dense matrix columns $N$. Density is defined as the number of non-zeros divided by the multiplication of the number of rows and cols for sparse matrix.}\label{fig:table}
\end{figure}

\textbf{Against the original TACO SpMM algorithms.} In this experiment, we compare the performance between TACO's original SpMM algorithms $\{<g\,nnz, c\,col>,1\}$ and $\{<x\,row,c\,col >,1\}$~\cite{senanayake:2020:scheduling} and two algorithms proposed by us, $\{<\frac{1}{g}\,row, c\,col>,r\}$ and $\{<1\,nnz , c\,col>,r\}$. We assign reasonable values to $g,c,x,$and $r$, and tune these parameters. We record the best performance of each algorithm on each dataset. From Table~\ref{tab-new} we conclude that {\newabs} brings 1.1x$\sim$1.2x normalized speedup. Fig.~\ref{fig:table} shows the detailed data.

\begin{table}[ht]
\begin{center}
\begin{minipage}{\textwidth}
\caption{Normalized performance of new algorithms}\label{tab-new}%
\begin{tabular*}{\textwidth}{@{\extracolsep{\fill}}lllll@{\extracolsep{\fill}}}
\toprule
 & RTX 3090  & RTX 2080 & Tesla V100 \\
\midrule
Speedup   & 1.191   & 1.098  & 1.223\\
\botrule
\end{tabular*}
\end{minipage}
\end{center}
\end{table}

\subsection{Generalization of {\framework}}
In this experiment, we implement our {\framework} to dgSPARSE library \footnote{\url{https://github.com/dgSPARSE}}, an open-source state-of-the-art SpMM and SDDMM library. We achieve up to 2.7x speedup on a certain SpMM algorithm.  
We keep the same sparse input matrix format (CSR) with dgSPARSE. After profiling, we find that row-major algorithms consistently outperform the col-major algorithms. Therefore, we target row-major. We are left with 4 algorithms: EB+SR+RM, EB+PR+RM, RB+SR+RM, RB+PR+RM. We will introduce the details of tuning RB+PR+RM and show the speedup.

To tune an actual GPU SpMM kernel, we require more fine-grained parameters than those in {\framework}. Parallelism is now two-fold: block-level and thread-level, instead of homogeneous threads. Besides, the memory hierarchy, such as the shared memory should be considered. Moreover, parallelism is limited in the physical world. For example, the largest thread-level parallelism is 1024 because a block has at most 1024 threads. The largest block-level parallelism is also finite(less than $2^{32}-1$). GridSize can be arbitrary because the extra blocks will be taken care of by GPU scheduler.

Tuning parameters for RB+PR+RM can be divided into two categories. The first is how many workers are assigned to process one chunk of data. The second is how many chunks of data are assigned to one worker. RB+PR+RM has 7 tunable parameters. A block process \textit{tileSz} real columns. \textit{workerSz} threads process one vectorized column and \textit{threadRw} sparse rows. \textit{groupSz} threads are synchronized. \textit{blockSz} denotes the number of threads per threadblock. \textit{workerDimR} denotes the block parallelism of sparse rows. A vectorized column has \textit{coarsenSz} consecutive real columns. If the overall sparse row parallelism is less than the number of rows in the sparse matrix, one thread may process more than one row. The tiling is``Dense major''; dense columns are fully parallelized. Specifically, \textit{blockDim.x = min(N, tileSz) / coarsenSz * workerSz}. Full source parallelism of one block is \textit{max(blockSz, blockDim.x * 2)}. In the dgSPARSE implementation, $tileSz=workerSz=groupSz=32$, workerDimR equals the number of rows of the sparse matrix , $threadRw=1$, $blockSz=256$, and \textit{coarsenSz=(N\%4==0)?4:(N\%2==0)?2:1}. 

Based on the insights of this paper, we should separate tiling and synchronization, add finer-grained parallelism, and more flexible workload of each thread. Therefore, we propose to tune four parameters: $<groupSz,blockSz,tileSz,workerDimR>$. Actually, workerDimR can be arbitrary. However, we set it to be power of 2 or reciprocal power of 2 times of the original value in order to explore the local area in the design space. As in {\framework} we set groupSz as 2,4,8,16, or 32. tileSz is power of 2 larger than groupSz, and depends on $N$. blockSz is set 128,256, or 512 which are common values for the number of threads per threadblock. We tune the RB+PR+RM kernel for $N=4,16,64,128$. From Table~\ref{tab-over-ori} we conclude that tuning can bring 1.6x$\sim$2.3x speedup over the original implementation\footnote{We open source our implementation at \url{https://github.com/dgSPARSE/dgSPARSE-Library/commit/9e3e4c18f40e76b97a805b8a9733258f7e9edeb6}.}.
\begin{table}[ht]
\begin{center}
\begin{minipage}{\textwidth}
\caption{Speedup over original implementation}\label{tab-over-ori}%
\begin{tabular*}{\textwidth}{@{\extracolsep{\fill}}lllll@{\extracolsep{\fill}}}
\toprule
Hardware & {geomean\footnotemark[1]}  & max & N \\
\midrule
\multirow{4}{*}{RTX 3090}& 2.295   & 4.316  & 128\\
                        & 2.181   & 4.432  & 64\\
                        & 1.997   & 4.271  & 16\\
                        & 2.046   & 7.819  & 4\\
\hline
\multirow{4}{*}{RTX 2080}& 1.938   & 4.379  & 128\\
                        & 1.927   & 4.430  & 64\\
                        & 1.995   & 5.019  & 16\\
                        & 2.307   & 8.582  & 4\\
\hline
\multirow{4}{*}{Tesla V100}   & 1.874   & 3.724  & 128\\
                        & 1.824   & 3.846  & 64\\
                        & 1.693   & 3.388  & 16\\
                        & 1.852   & 6.114  & 4\\
\botrule
\end{tabular*}
\footnotetext[1]{We use geometric mean to reduce outlier bias.}
\end{minipage}
\end{center}
\end{table}

Because DA-SpMM introduces a decision tree model to choose the best configuration for a given sparse matrix, we further explore the maximum speedup that dynamic choices can bring. This experiment examines the necessity of designing a new model to choose the best parameters. From Table~\ref{tab-over-static} we conclude that the most significant speedup of dynamic choices is 1.1x$\sim$1.4x.
\begin{table}[ht]
\begin{center}
\begin{minipage}{\textwidth}
\caption{Speedup over static implementation}\label{tab-over-static}%
\begin{tabular*}{\textwidth}{@{\extracolsep{\fill}}lllll@{\extracolsep{\fill}}}
\toprule
Hardware & geomean  & N & Best static\\
\midrule
\multirow{4}{*}{RTX 3090}& 1.124   & 128 & $<8,256,8,1/2>$\\
                        & 1.114   & 64  & $<4,256,8,1/2>$\\
                        & 1.310   & 16  & $<8,256,8,1/2>$\\
                        & 1.406   & 4  & $<8,256,8,1>$\\
\hline
\multirow{4}{*}{RTX 2080}& 1.095   & 128  & $<4,256,8,1/2>$\\
                        & 1.114   & 64  & $<4,256,8,1/2>$\\
                        & 1.276   & 16  & $<4,256,8,1/2>$\\
                        & 1.310   & 4  & $<4,256,8,1/2>$\\
\hline
\multirow{4}{*}{Tesla V100}   & 1.137   & 128  & $<8,256,8,1/2>$\\
                        & 1.177   & 64  & $<8,256,8,1/2>$\\
                        & 1.367   & 16  & $<8,256,8,1>$\\
                        & 1.326   & 4  & $<8,256,8,1>$\\
\botrule
\end{tabular*}
\end{minipage}
\end{center}
\end{table}
\section{Conclusion}\label{sec:conclusion}
We propose {\framework} to analyze sparse-dense hybrid algebra and propose new SpMM designs. Based on {\framework} propose a new abstraction {\newabs} to sparse compilers and remedy the missing optimization opportunities. First, we implement the new abstraction in TACO and achieve up to 1.2x speedup over TACO's original SpMM kernels. Then, we use {\framework} to tune an SpMM algorithm in dgSPARSE and get 1.6x$\sim$2.3x speedup on the tuned algorithm. In the future, {\framework} can be exposed as an auto-tuning API for users to explore different synchronization granularity and reduction strategy for sparse-dense hybrid algebra.



\end{document}